\documentstyle[12pt,epsfig,psfig]{article}

\newcommand{\newc}{\newcommand}
\newc{\gsim}{\lower.7ex\hbox{$\;\stackrel{\textstyle>}{\sim}\;$}}
\newc{\lsim}{\lower.7ex\hbox{$\;\stackrel{\textstyle<}{\sim}\;$}}
\newc{\gev}{\,{\rm GeV}}
\newc{\mev}{\,{\rm MeV}}
\newc{\ev}{\,{\rm eV}}
\newc{\kev}{\,{\rm keV}}
\newc{\tev}{\,{\rm TeV}}
\def\ln{\mathop{\rm ln}}

\newc{\mz}{M_Z}
\newc{\mpl}{M_*}
\newc{\mw}{m_{\rm weak}}
\def\beq{\begin{equation}}
\def\eeq{\end{equation}}
\def\bea{\begin{eqnarray}}
\def\eea{\end{eqnarray}}
\newc{\ie}{{\it i.e.}}          \newc{\etal}{{\it et al.}}
\newc{\eg}{{\it e.g.}}          \newc{\etc}{{\it etc.}}
\newc{\cf}{{\it c.f.}}
\def\bar#1{\overline{#1}}

\def\inv{^{\raise.15ex\hbox{${\scriptscriptstyle -}$}\kern-.05em 1}}
\def\lbar{{\lower.35ex\hbox{$\mathchar'26$}\mkern-10mu\lambda}} 

\def\om#1#2{\omega^{#1}{}_{#2}}

\def\to{\rightarrow}

\let\<=\langle
\let\>=\rangle

\let\+=\uparrow

\let\Ga=\Gamma

\let\si=\sigma

\let\om=\omega

\begin{document}
\thispagestyle{empty}
\vspace*{.5cm}
\noindent
\hspace*{\fill}{\large CERN-TH/2002-363}\\
\hspace*{\fill}{\large OUTP 02 45P}\\
\vspace*{2.0cm}

\begin{center}
{\Large\bf Calculable~Corrections~to~Brane~Black~Hole~Decay~II:}\\[.2cm]
{\Large\bf Greybody Factors for Spin 1/2 and 1}
\\[2.5cm]
{\large Panagiota Kanti${}^*$ and John March-Russell${}^{*\dagger}$
}\\[.5cm]
{\it ${}^*$Theory Division, CERN, CH-1211 Geneva 23, Switzerland}
\\[.2cm]
{\it ${}^{\dagger}$Theoretical Physics, Oxford University, 1 Keble Rd., OX1
3NP, UK}
\\[.2cm]
(March, 2003)
\\[1.1cm]

{\bf Abstract}\end{center}
\noindent
The production of black holes in extra-dimensional brane-world
theories can lead to detectable signals via the Hawking evaporation of the
black hole to brane-localised Standard Model modes. We calculate, as a
function of partial wave number and number of toroidally compactified extra
dimensions, the leading correction to the energy spectrum of such
Hawking radiation (the greybody factors) for decay into spin 1/2
fermions and spin 1 gauge fields localised on the Standard Model
brane.  We derive the associated improved differential emission rates
for both types of fields.  We provide both simple expressions 
for the leading behavior of the greybody factors in the low-energy
limit $\om r_H \ll 1$, and numerical evaluation of our full analytic
expressions for the emission rates, valid for $\om r_H \sim 1$.  The
full analytic expressions demonstrate that both the greybody factors and
emission rates are enhanced as the number of extra dimensions increases.

\newpage

\setcounter{page}{1}

\section{Introduction}

One of the most spectacular consequences of brane-world
theories~\cite{ADD,early,RS} with low fundamental gravitational scale,
$\mpl\gsim\tev$ (for some works on the structure of these theories, on their
cosmological implications and experimental signatures, see \cite{activity}),
is the modification of the properties of small black holes~\cite{admr} that,
in principle, allows them to be copiously produced at the LHC and in other
high-energy processes~\cite{bf,gt,dl,support,voloshin,bhphenom,cosmic}.
In a collider setting, the dominant signal for such black hole production
arises from the Hawking evaporation products of the black hole, in particular
the characteristic spectrum of such radiation.
In a previous paper \cite{KMR1}, we considered the leading semiclassical 
corrections to the spectrum of Hawking radiation for a higher-dimensional
black hole evaporating to scalar modes.  We considered both a scalar field
localised to the brane as well as the situation in which it
could propagate in the full $(4+n)$-dimensional bulk, and computed the
corresponding greybody factors as functions of energy $\om$, angular momentum
$j$, and number of (flat) extra dimensions $n$.  The scalar case
is both a useful testing place of the methods used in such a computation, and
potentially relevant for Higgs, or other scalar boson production from
black holes formed in collisions at the LHC or other high energy colliders. However,
from a practical point of view the most important cases to consider are those
of the production of SM fermions or SM gauge bosons to which we turn in this paper.

In both cases the 
fields that need to be considered are brane localised.  For fermions
this is because of the chiral nature of the SM fermions, since a fermion
propagating in the bulk is necessarily non-chiral from the 4d
perspective.\footnote{The RHD neutrino can propagate in the bulk,
leading to an alternate, non-seesaw explanation of the lightness of the observed
neutrino states~\cite{AHDDMR}. Thus in principle it is possible to consider the
Hawking decay of a black hole to these bulk neutrino states.  However, from
the perspective of the
brane observer this leads only to a missing energy signal which is not
distinguishable from the many other sources of missing energy in brane-world
theories.  Therefore we do not consider this possibility in the present work.}   
It is possible to consider SM gauge fields propagating in the
higher-dimensional bulk, or a subspace thereof \cite{DDG}.  However,
the already existing collider constraints bound the size of such gauge extra
dimensions to $R_{\rm gauge}^{-1} \gsim 3\tev$, while, in the best
possible case where only the graviton propagates in the extra
dimensions, the limits on
the new fundamental higher-dimensional Planck scale are of
order $\mpl\gsim 1.2\tev$~\cite{pdg}.   On the other hand,
the validity of the semiclassical description of black holes requires that
their horizon radius $r_H$ be large compared to the higher dimensional
Planck length $r_H \gg 1/\mpl$, which given the just cited bounds, as
well as the limits on the maximum mass of black hole that can be produced at
the LHC, implies that $r_H$ must be somewhat large compared to $R_{\rm
gauge}$, too.  In this case the gauge fields can also be considered as
effectively brane localised, the `brane' in question having some non-trivial
substructure. To leading order this substructure does not change the greybody
factors we compute for the spin 1 case.

Note that the properties of a black hole, in particular its
Schwarzschild radius, $r_H$, and production cross section, differ from
the 4d result provided that $r_H$ is smaller than the size of at least
some of the extra dimensions $r_H < R$.  If the brane tension does not
strongly perturb the $(4+n)$-dimensional black hole solution, and the 
extra dimensional space is approximately flat on the scale $r_H$, then
the black hole is well-described as a angularly
symmetric $(4+n)$-dimensional black hole centered on the brane,
but extending out into the $n$ extra dimensions, leading to the
expression~\cite{admr,mp},
\beq
r_H = {1\over\mpl} \left(M\over\mpl\right)^{1\over n+1}
\left(8 \Gamma((n+3)/2)\over (n+2) \pi^{(n+1)/2}\right)^{1/(n+1)}\,,
\label{eq:rh}
\eeq
for the Schwarzschild radius.  Here, we follow these commonly made
assumptions, which have the consequence of limiting our discussion
to the Hawking decay of black holes in the context of
Arkani-Hamed-Dimopoulos-Dvali like theories~\cite{ADD}.  In
particular, black holes in theories of
the Randall-Sundrum type \cite{RS} require a more involved treatment (for a
recent paper considering the decay of RS black holes see~\cite{EGBK}).

After such black holes are produced, they decay by the emission
of Hawking radiation, and it is expected that they will decay mostly
to particles on our brane~\cite{ehm}. The spectrum of radiation,
with a temperature $T_{BH}= (1 + n)/( 4 \pi  r_H)$ (in $G = k_B =
c =\hbar=1$ units), is given by the Hawking formula~\cite{hawking}
\beq
\frac{dE(\om)}{dt} = \sum_{j,b} \sigma_{j,b}(\om)\,{\om  \over
\exp\left(\om/T_{BH}\right) \mp 1}\,\frac{d^{n+3}k}{(2\pi)^{n+3}}\,,
\label{eq:greybody}
\eeq
where $j$ labels the total angular momentum quantum number,
$b$ labels any other quantum numbers of the emitted
particle, as well as the particle type, and in the phase-space
integral $|k| = \om$ for a massless particle.  Here
$\si(\om)$, known as the `greybody factor', is an energy-dependent 
function arising from the backscattering of part of the outgoing radiation
back into the black hole by the non-trivial metric in the region
exterior to the black hole horizon (for early works on the
emission spectra of 4d black holes, see \cite{classics}).

In this paper we calculate the greybody factors 
and emission rates for spin 1/2 and spin
1 particles localised to the SM brane in the presence of a
non-rotating black hole of the type described above.  It is expected that
such a non-rotating black hole is a good approximation to the exact
solution after the initial `balding' and `spin-down' phases~\cite{gt}.
As discussed for the scalar case in our first paper \cite{KMR1}, the
greybody factor is computed using its equality with
the absorption cross section for the appropriate type of particle
incident on the background metric that describes the brane black hole.
The equality of greybody factors to absorption cross sections
implies that the greybody factors do not invalidate the thermal nature of the
black hole. 

Finally, it is important to recall that the semiclassical
calculation of Hawking emission is only reliable when
the energy of the emitted particle is small compared to the black
hole mass $\om \ll M$, since only in this case is it correct
to neglect the back reaction of the metric during the emission process.
This in turn requires that the Hawking temperature obeys the relation
$T_{BH}\ll M$, which
is equivalent to demanding that the black hole mass $M\gg \mpl$.
Inevitably this condition
breaks down during the final stages of the decay process, but for those
black holes of initial mass larger than $\mpl$ most of the evaporation
process is within the semi-classical regime.

In Section~2 of this paper, we present the general equation for the radial
part of the wave function of fields with spin 1/2 and 1, as well as 
a discussion of the metric used in calculating the greybody factors.
We proceed to solve this equation at the asymptotic regimes of the
near-horizon and far-field zones, for arbitrary number of extra dimensions
and angular momentum number, and we match them in an intermediate
zone in order to construct a complete, smooth solution over the whole radial
regime. 

In section~3, the greybody factors for fermionic and gauge
fields are computed.  Our full analytic results are contained in
Eqs.~(35-38), (40), (46), and (49).  Although analytically
complicated, these full results are necessary for the accurate
determination of the Hawking emission spectrum from brane black holes
(see section 4).  We expect they will be of use to those searching
for black holes at
current and future colliders, or in ultra-high-energy-cosmic-ray
collisions.  However, for ease of comparison with previous
studies of greybody factors we derive simplified expressions for
the leading corrections to the spectrum of the Hawking radiation
valid for $\omega r_H  \ll 1$.  These expressions
are presented in Eq. (\ref{fermionsi}) for
spin 1/2 and Eq. (\ref{gaugesi}) for spin 1, for arbitrary $j$ and $n$.  
Their values for $n=2,4,6$ extra dimensions,
and for the first three, and thus most important, angular momentum modes
(with $j =\frac{1}{2}$, $\frac{3}{2}$, $\frac{5}{2}$  for spin 1/2,
and $j=$1, 2, 3 for spin 1) are tabulated in Table~1 and Table~2,
respectively. 

In section 4, we proceed to calculate the full greybody factors 
and emission rates for spin 1/2 and spin 1 fields by a numerical
evaluation of our full analytic results of sections 2 and 3 (valid
beyond the simplified limit $\om r_H \ll 1$).
A similar analysis for spin 0 fields, based on the results
derived in \cite{KMR1}, is also included in this section.
The results are presented in figures 1,2 and 3.
Our conclusions are summarized in section 5.


\section{General equation for spin 1/2 and 1 fields and its solution}

The purpose of this paper is to calculate
the greybody factors for the emission of fermion and gauge fields in 
the background of the metric on their 4-dimensional brane induced from
a $(4+n)$-dimensional Schwarzschild-like black hole centered on the brane. 
The starting point of our analysis will be the general equation for the
propagation of fields with spin $s=1/2$ and 1 in an arbitrary 4-dimensional
spherically-symmetric background.   Using the Newman-Penrose formalism
\cite{chandra} (see \cite{CL} for a compact review), this general
field equation may be written as
\beq
\nabla_{AA'}\,\psi^{AB_1 ... B_{2s-1}}=0 .
\eeq
Following the analysis of Cvetic and Larsen \cite{CL}, it is possible
to derive from this general equation a set of $2s$
differential equations for the components of the wave function of the
spin $s$ field.  Specifically,
assuming a general, spherically-symmetric background of the form
\beq
ds^2 = - \frac{\Delta}{\Sigma}\,dt^2 +\frac{\Sigma}{\Delta}\,dr^2 + 
\Sigma^2\,(d\theta^2+ \sin^2\theta\,d\varphi^2)\,,
\label{general}
\eeq
and using a factorized ansatz for each component of the wave function of the
field of the form
\beq
\psi_{k-s}= \Delta^{-k/2}\,(2\Sigma)^{k/2-s}\,R_{k-s}(r)\,S_{k-s}(\Omega)\,,
\eeq
the set of coupled differential equations separates into two
sets of decoupled 
equations for the radial and angular parts of the wave function. In the above
expression, $k=0, ..., 2s$ labels the helicity of the components, and 
$S_{k-s}(\Omega)=e^{i m \varphi}\,d^{(j)}_{km}(\theta)$, where $j$ and $m$
are the total angular momentum number and its projection on a fixed axis,
respectively, and $d^{(j)}_{km}$ stand for the rotation matrices.
Focusing on the radial part of the wave function, the form of
the corresponding equation turns out to depend on the helicity of the
component. For example, for the upper component $\psi_s$ ($k=2s$), it
takes the form~\cite{CL}:
\bea
&~& \hspace*{-1.5cm}
\Delta^s\,\frac{d \,}{dr}\,\biggl[ \Delta^{1-s}\,\frac{d R_s}{dr}\biggr] +
\biggl\{\frac{\Sigma^2 \om^2 -i s\,\om\,\Sigma\,\partial_r\Delta}{\Delta}
+ 2 i s\,\om\,\partial_r \Sigma -\Lambda + \nonumber \\[3mm]
&~& \hspace*{0.5cm}\Delta\,\Bigl(s-\frac{1}{2}\Bigl)\,\biggl[\partial_r\,
\biggl(\frac{\partial_r \Sigma}{\Sigma}\biggr) + 
\Bigl(s-\frac{1}{2}\Bigl)\,\biggl(\frac{\partial_r \Sigma}{\Sigma}\biggr)^2
+ (1-s)\,\frac{\partial_r \Sigma}{\Sigma}\,\frac{\partial_r \Delta}{\Delta}
\biggr]\,\biggr\}\,R_s=0\,.
\label{master}
\eea
In the above, $\Lambda\equiv j\,(j+1)-s\,(s-1)$. As we will shortly
see, solving
the above equation for the upper component of the field only is
sufficient to lead to the determination of the absorption cross
section, and thus the greybody factor.

In the presence of additional space-like dimensions, the higher-dimensional 
spherically symmetric generalizations of the 4-dimensional
Schwarzschild solution have a line element given by
\beq
ds^2 = - h(r)\,dt^2 +h(r)^{-1}\,dr^2 + r^2\,d\Omega_{2+n}^2,
\label{bhmetric}
\eeq
where
\beq
h(r) = 1-\biggl(\frac{r_H}{r}\biggr)^{n+1}\,.
\label{h-fun}
\eeq
The angular part of the above higher-dimensional metric tensor is
\beq
d\Omega_{2+n}^2=d\theta^2_{n+1} + \sin^2\theta_{n+1} \,\biggl(d\theta_n^2 +
\sin^2\theta_n\,\Bigl(\,... + \sin^2\theta_2\,(d\theta_1^2 + \sin^2 \theta_1
\,d\varphi^2)\,...\,\Bigr)\biggr)\,,
\eeq
with $0 < \varphi < 2\pi$ and $0<\theta_i<\pi$, for $i=1,...,n+1$.
It is easy to check that, at distances $r\gg R$, where $R$ the size of the extra
compact dimensions, the above metric tensor reduces to the usual 4-dimensional
Schwarzschild solution.

Under the assumption that the Standard Model fields, both fermions and gauge bosons,
are restricted to live on a 4-dimensional brane, they propagate in a
gravitational background which is given by the induced 4d metric on
the brane following from the higher-dimensional
black-hole background~(\ref{bhmetric}). This follows if we set
$\theta_n=\pi/2$, for $n \ge 2$.  Specifically it takes the form
\beq
ds^2 = - h(r)\,dt^2 +h(r)^{-1}\,dr^2 + r^2\,d\Omega_2^2\,.
\label{projected}
\eeq
Comparing Eqs. (\ref{general}) and (\ref{projected}), we may easily see that the
general metric functions appearing in Eq. (\ref{general}) are now given by:
$\Sigma=r^2$ and $\Delta=h r^2$. Substituting in the general, radial differential
equation (\ref{master}), we obtain the following simplified equation:
\bea
&~& \hspace*{-1.5cm}
(h r^2)^s\,\frac{d \,}{dr}\,\biggl[ (h r^2)^{1-s}\,\frac{d R_s}{dr}\biggr] +
\biggl\{\frac{\om^2 r^2}{h} + 2i s\,\om\,r -\frac{i s \om\,r}{h}\,
(n+1)\,(1-h) - \nonumber\\[3mm]
&~& \hspace*{4.8cm}- \Lambda - (2s-1)\,(s-1)\,(n+1)\,(1-h)\biggr\}\,R_s=0\,.
\label{master1}
\eea
In the above, we have used the relation
\beq
\frac{dh}{dr}=\frac{(n+1)}{r}\,(1-h)\,,\label{dh}
\eeq
that follows from the definition Eq.~(\ref{h-fun}).
In this paper, we are going to study the cases of fermion ($s=1/2$) and
gauge-boson fields ($s=1$). For those fields, the last term appearing
inside the curly brackets in Eq. (\ref{master1}) vanishes independently
of the number of projected extra dimensions and, therefore, it will be
ignored in the following analysis.

In order to compute the greybody factor for the Hawking radiation in the 
gravitational background described in Eq. (\ref{projected}), through
the emission of fermion and gauge-boson fields,  we need to solve
Eq. (\ref{master1}) for the radial
part of the emitted field in the region outside the horizon of the black hole 
and all the way to infinity. Due to the complexity of the differential
equation, the exact solution appears impossible to derive. As in \cite{method}
\cite{KMR1}, (see also \cite{others} for related works) we are going
to follow an approximation method that is
suitable for low energies $\om r_H \ll 1$. The method involves solving
Eq. (\ref{master1}) first in the vicinity of the black hole, then
at infinite distance from it and, finally, matching the two asymptotic
solutions in an intermediate regime. This method will help us derive
the form of the absorption coefficient in the equivalent scattering 
problem of an incoming wave propagating in the background (\ref{projected}),
in terms of which the absorption cross section, and thus the greybody factor
of the emitted radiation is defined.

As we will shortly see, different components of the emitted field carry
different parts of the particle wave function. For an emitted field with spin
$s \neq 0$, it is only the upper, $\psi_s$, and lower, $\psi_{-s}$, components
that are the radiative ones. The upper component will turn out to correspond
mainly to the incoming wave while the lower component corresponds mainly
to the reflected wave. Moreover, the
equation for the lower component is the charge conjugate of the one for the
upper component, therefore solving for any one of them provides the solution
for the other one. Here, we will concentrate on the incoming wave and
compute the absorption coefficient by comparing the incoming flux at 
the asymptotic regimes of the horizon and infinity. The matching of
the two solutions in the intermediate regime, however, is still
necessary.  We impose the following normalizations for the incoming mode
of a field, with spin $s$,
\beq
R^{(h)}_s = A^{(h)}_s\,h^{-i \beta_{BH} \om/4\pi},
\label{norm-h}
\eeq
at the horizon of the black hole, where $\beta_{BH}=1/T_{BH}$, and
\beq
R^{(\infty)}_s = A^{(\infty)}_s\,(2 \om r)^{2s-1}\,e^{-i \om r}
\label{norm-ff}
\eeq
at infinity. In the forthcoming sections, and for the solution of the general,
radial equation (\ref{master1}), we closely follow the analysis presented
in Ref. \cite{KMR1} for the case of the emission of scalar fields.

\subsection{Solving the general equation in the near-horizon zone}

In the vicinity of the black hole, Eq. (\ref{master1}) can be solved by making 
a change of variable, $r \to h$.  Using Eqs. (\ref{h-fun}) and 
(\ref{dh}), we may write the general-spin field equation in the form
\bea
&~& \hspace*{-2cm} 
h\,(1-h)\,\frac{d^2 R}{dh^2} + \biggl[(1-s)\,(1-h) -\frac{(n+2s)}{(n+1)}
\,h\biggr]\,\frac{d R}{dh} + \nonumber \\[2mm]
&~& \hspace*{2cm} \biggl[\,\frac{(\om r_H)^2}{(n+1)^2 h (1-h)} 
+ \frac{2is\,\om r_H -\Lambda}{(n+1)^2 (1-h)}-
\frac{i s\,\om r_H}{(n+1)\,h}\,\biggr] R=0\,,
\label{NH-1}
\eea
where, for simplicity of the notation, the subscript $s$ has been dropped
from $R$. By using the redefinition  $R(h)=h^\alpha (1-h)^\beta F(h)$, the
above equation takes the form of a hypergeometric equation
\beq
h\,(1-h)\,\frac{d^2 F}{dh^2} + [c-(1+a+b)\,h]\,\frac{d F}{dh} -ab\,F=0\,,
\label{hyper}
\eeq
with
\beq
a=\alpha + \beta +\frac{s + n\,(1-s)}{(n+1)}\,, \qquad 
b=\alpha + \beta\,, \qquad c=1-s + 2 \alpha\,.
\eeq
The power coefficients $\alpha$ and $\beta$, in turn, are found by solving 
second-order algebraic equations leading to the results 
\beq
\alpha_{+} = s+\frac{i \om r_H}{n+1}\,, \qquad
\alpha_{-} = -\frac{i \om r_H}{n+1}\,, 
\eeq
and
\beq
\beta_{\pm} =\frac{1}{2 (n+1)}\,\biggl[\,1-2s \pm \sqrt{(1+2j)^2 -
4 \om^2 r_H^2 -8 is \om r_H}\,\,\biggr]\,,
\label{al-be}
\eeq\\
%
respectively. The general solution of the hypergeometric equation (\ref{hyper}) is
\medskip
\beq
R_{NH}(h)=A_- h^{\alpha}\,(1-h)^\beta\,F(a,b,c;h) +
A_+\,h^{-\alpha}\,(1-h)^\beta\,F(a-c+1,b-c+1,2-c;h)\,,
\label{NH-gen}
\eeq

\noindent
where $A_{\pm}$ are arbitrary constants. Expanding the above solution in the limit
$r \rightarrow r_H$ (equivalently $h \rightarrow 0$) we obtain the asymptotic behaviour
\smallskip 
\beq
R_{NH} \simeq A_-\,h^\alpha + A_+\,h^{-\alpha} = 
A_-\,h^s\,\exp\Bigl(i \om r_H^{n+2} y\Bigr) + 
A_+\,h^{-s}\,\exp\Bigl(-i \om r_H^{n+2} y\Bigr)
\label{NH-alpha+}
\eeq
for $\alpha=\alpha_+$, and 
\beq
R_{NH} \simeq 
A_-\,\exp\Bigl(-i \om r_H^{n+2} y\Bigr) + A_+\,\exp\Bigl(i \om r_H^{n+2} y\Bigr)
\label{NH-alpha-}
\eeq
for $\alpha=\alpha_-$. In the above, we have used the `tortoise' coordinate $y$
defined as
\beq
y=\frac{\ln h(r)}{r_H^{n+1}\,(n+1)}\,.
\label{ycoord}
\eeq
The choice $\alpha=\alpha_+$ leads to a solution with an outgoing wave of zero
amplitude at the horizon, and an incoming wave with infinite amplitude. This
is an irregular solution which must be discarded. On the other hand, the
choice of $\alpha=\alpha_-$ leads to regular incoming and outgoing waves
with amplitude unity at the horizon. The boundary condition that only incoming
modes are to be found in the region outside the horizon of a black hole 
forces us also to set $A_+=0$. After these choices are made, the solution
near the horizon has exactly the normalization of the fields
described in Eq. (\ref{norm-h}), since $\alpha_-=- i \om\,\beta_{BH}/4\pi$,
thus determining the asymptotic normalization constant to be $A^{(h)}_s=A_-$.
Turning to the $\beta$ coefficient, the criterion for the convergence
of the hypergeometric function $F(a,b,c;h)$
\beq
{\bf Re}\,(c-a-b) = -\biggl(\,\pm \frac{1}{n+1}\,\sqrt{(1+2j)^2 -
4 \om^2 r_H^2 -8 is \om r_H}\,\biggr) >0 
\eeq
clearly demands that we choose $\beta=\beta_{-}$.

We finally need to shift the hypergeometric function towards large values of $r$.
This may be done, by using a standard linear transformation formula \cite{abram},
in the following way:
\bea
&~&  \hspace*{-0.9cm}
R_{NH}(h)=A_- \,h^{\alpha} \biggl[(1-h)^\beta \frac{\Gamma(1-s + 2\alpha)
\Gamma\Bigl(-2\beta+\frac{1-2s}{n+1}\Bigr)}{\Gamma(\alpha-\beta +1-s)
\Gamma\Bigl(\alpha-\beta+\frac{1-2s}{n+1}\Bigr)}
F(a,b,a+b-c+1;1-h)\nonumber\\[2mm]
&~&
\hspace*{-1.05cm}+(1-h)^{-\beta+(1-2s)/(n+1)}\frac{\Gamma(1-s+2\alpha)
\Gamma\Bigl(2\beta-\frac{1-2s}{n+1}\Bigr)}
{\Gamma(\alpha+\beta) \Gamma\Bigl(\alpha+\beta+\frac{s+n\,(1-s)}{n+1}\Bigr)}
F(c-a,c-b,c-a-b+1;1-h)\biggr].\nonumber\\
\eea
We can now ``stretch" the above expression towards the intermediate regime, by
expanding in the limit $r \rightarrow \infty$, or $h \rightarrow 1$, and take
\bea
R_{NH}(h) &\simeq& A_-\,\Bigl(\frac{r_H}{r}\Bigr)^{-(j+s)}\,
\frac{\Gamma(1-s+2\alpha)\,
\Gamma\Bigl(-2\beta+\frac{1-2s}{n+1}\Bigr)}{\Gamma(\alpha-\beta+1-s)\,
\Gamma\Bigl(\alpha-\beta+\frac{1-2s}{n+1}\Bigr)}\nonumber \\[2mm]
&+& A_-\,\Bigl(\frac{r_H}{r}\Bigr)^{j-s+1}
\,\frac{\Gamma(1-s+2\alpha)\,\Gamma\Bigl(2\beta-\frac{1-2s}{n+1}\Bigr)}
{\Gamma(\alpha+\beta)\,\Gamma\Bigl(\alpha+\beta+\frac{s+n\,(1-s)}{n+1}\Bigr)}\,.
\label{NH-large}
\eea
Note that, in order to simplify the procedure of the ``matching", the low
energy limit has been taken in the expression of the $\beta$ coefficient in
the power of $r$. No expansion has been made, up to this point, in the
arguments of the Gamma functions.

\subsection{Solving the general equation in the far-field zone}

We now need to find the far-field zone
solution before being able to match the two asymptotics in the intermediate
regime. Going back to the general equation (\ref{master1}), and in
the limit $r \rightarrow \infty$ or $h \rightarrow 1$, we obtain
\beq
\frac{d^2 R}{d r^2} + \frac{2 (1-s)}{r}\,\frac{d R}{d r} +
\biggl(\om^2 +\frac{2is \om}{r} -\frac{\Lambda}{r^2} \biggr)\,R=0\,.
\eeq
In order to solve the above equation, we set: $R=e^{-i \om r} \,r^{j+s}\,\tilde R(r)$.
By also making a change of variable $z=2 i \om r$, the above equation adopts the
form of a confluent hypergeometric equation
\beq
z\,\frac{d^2 \tilde R}{d z^2} + (b- z)\,\frac{d \tilde R}{d z} - a \tilde R=0\,,
\eeq
with $a=j-s+1$ and $b=2j+2$, and general solution
\beq
\tilde R(z)=B_+\,M(a,b,z) + B_-\,U(a, b, z)\,,
\eeq
where $M$ and $U$ are the Kummer functions, and $B_\pm$ are arbitrary coefficients.
We may therefore write the complete solution for the radial function at infinity as
\beq
R_{FF}(r) = e^{-i \om r} \,r^{j+s}\,\Bigl[B_+\,M(j-s+1,\,2j+2,\,2i\om r) 
+ B_-\,U(j-s+1,\,2j+2,\,2 i \om r)\Bigr]\,.
\label{FF-sol}
\eeq

We first need to expand the above expression for large values of $r$. This
expansion will help us make sure that the emitted fields have the far-field
normalization given in Eq. (\ref{norm-ff}) and will enable us to compute
the incoming flux at infinity. In the limit $r \rightarrow \infty$, we find
\beq
R_{FF}(r) = \frac{e^{-i \om r} \,r^{2s-1}}{(2 i \om)^{j-s+1}}\,
\biggr[B_- + \frac{B_+\,e^{i\pi(j-s+1)}\,\Gamma(2j +2)}{\Gamma(j+s+1)}\biggr]
+ \frac{e^{i \om r} \,B_+\,\Gamma(2j+2)}{\Gamma(j-s+1)\,(2i\om)^{j+s+1}\,r} 
+ ... 
\eeq
The first part of the above expression gives the incoming wave at
infinity, and thus, by comparison with Eq.~(\ref{norm-ff}), defines the
coefficient $A^{(\infty)}_s$.  The second part gives the outgoing
wave.  This outgoing part is suppressed as it should be. We
remind the reader that the above solution corresponds only to the upper
component of the emitted field (either to $\psi_{1/2}$ or to $\psi_1$)
which in turn corresponds to the incoming wave. The charge conjugate of
the above solution gives the lower component (the other radiative
component) of the emitted field, which will have a dominant outgoing
wave and a suppressed incoming one.  

We may now stretch the far-field solution (\ref{FF-sol}) towards small
values of $r$. In the limit $\om r \ll 1$, we therefore obtain
\beq
R_{FF}(r)= B_+\,r^{j+s} + \frac{B_-}{r^{j-s+1}}\,
\frac{\Gamma(2j+1)}{\Gamma(j-s+1)\,(2 i \om)^{2j+1}}\,.
\label{FF-small}
\eeq
Matching the two solutions (\ref{NH-large}) and (\ref{FF-small}) in the
intermediate regime, we obtain the relations
\bea
B_+ &=& \frac{A_-}{r_H^{j+s}}\,\frac{\Gamma(1-s+2\alpha)\,
\Gamma\Bigl(-2\beta+\frac{1-2s}{n+1}\Bigr)}{\Gamma(\alpha-\beta+1-s)\,
\Gamma\Bigl(\alpha-\beta+\frac{1-2s}{n+1}\Bigr)}\,,\label{B+}\\[2mm]
B_- &=& \frac{A_-\,r_H^{j-s+1}\,(2 i \om)^{2j+1}\,\Gamma(1-s+2\alpha)\,
\Gamma\Bigl(2\beta-\frac{1-2s}{n+1}\Bigr)\,\Gamma(j-s+1)}
{\Gamma(\alpha+\beta)\,\Gamma\Bigl(\alpha+\beta+\frac{s+n\,(1-s)}{n+1}\Bigr)
\,\Gamma(2j+1)}\,. \label{B-}
\eea
With the determination of the above arbitrary coefficients, we have now
completed the determination of the complete solution for both fermions and
gauge-boson fields propagating in the background of a projected 
$(4+n)$-dimensional Schwarzschild black hole on a 4-dimensional brane.
We now proceed to the calculation of the greybody factor in each case
for an arbitrary number of extra compact space-like dimensions.

\section{Greybody factors for the emission of fermionic and gauge fields}

As we will shortly see, the expression of the greybody factors for Hawking
radiation for fermions and gauge fields, involves the quantity
$|A^{(h)}/A^{(\infty)}|^2$ determined by the two normalization coefficients
appearing in the asymptotic solutions (\ref{norm-h}) and (\ref{norm-ff}).
By using the relations (\ref{B+})-(\ref{B-}), this takes the form
\beq
\frac{A^{(h)}}{A^{(\infty)}} =\frac{A_- \,(2\om)^{j+s}\,e^{i\pi\,(j-s+1)/2}}
{\biggr[B_- + \frac{\textstyle B_+\,e^{i\pi\,(j-s+1)}\,\Gamma(2j +2)}
{\textstyle \Gamma(j+s+1)}\biggr]}=
\frac{(2\om r_H)^{j+s}\, e^{-i \pi\,(j-s+1)/2}}
{\Gamma(1-s+2 \alpha)\,\Bigr[\,C\,(\om r_H)^{2j+1} + D\,\Bigr]}\,,
\eeq
where the coefficients $C$ and $D$ stand for
\bea
C &=& \frac{2^{2j+1}\,e^{i\pi\,(s-1/2)}\,\Gamma\Bigl(2\beta-\frac{1-2s}{n+1}\Bigr)\,
\Gamma(j-s+1)}{\Gamma(\alpha+\beta)\,\Gamma\Bigl(\alpha+\beta+\frac{s+n\,(1-s)}{n+1}\Bigr)
\,\Gamma(2j+1)}\,, \label{C}\\[2mm]
D &=& \frac{\Gamma(2j+2)\,\Gamma\Bigl(-2\beta+\frac{1-2s}{n+1}\Bigr)}
{\Gamma(\alpha-\beta+1-s)\,\Gamma\Bigl(\alpha-\beta+\frac{1-2s}{n+1}\Bigr)
\,\Gamma(j+s+1)}\,.\label{D}
\eea
The measure squared of the ratio $A^{(h)}/A^{(\infty)}$ can then be written as
\beq
\biggl|\frac{A^{(h)}}{A^{(\infty)}}\biggl|^2 =
\frac{(2\om r_H)^{2(j+s)}} {|\Gamma(1-s+2 \alpha)|^2\,
\Bigr[\,C C^*(\om r_H)^{4j+2} + (\om r_H)^{2j+1}\,(C D^* + C^* D)+
D D^*\,\Bigr]}\,.
\label{ratio1}
\eeq
The derivation of a final explicit result for the greybody factors or
emission rates from the above expression requires the evaluation of
Gamma functions for complex arguments. This task is a tedious
procedure with the result being highly complicated and unilluminating.
Accurate evaluations of the greybody factors and associated
emission rates do require use of the above expressions, and so
in section 4, we present numerical results for the final expression
of the greybody factors, as a function of $\omega r_H$, by using the exact
value of the above ratio. 
However, {\it for comparison with previous studies}, 
and as in the case of the analysis for the emission of scalar fields
\cite{KMR1}, a simplified expression may be derived in the low-energy limit
$\om r_H \ll 1$. Expanding the Gamma functions in this limit, we may express
the denominator of the fraction appearing in Eq. (\ref{ratio1}) in a
power series in $(\om r_H)$ keeping only the leading term. The
result of the expansion
depends strongly on the spin of the emitted particle, therefore, we will now
discuss separately the results for each case.  We emphasise that
applications in which accurate values of the emissions rate are
required should {\it not} use these simplified expressions valid 
only for $\om r_H \ll 1$, but rather should use the full semiclassical
results derived above (together with Eqs.~(40), (46), and (49))
as in the evaluation presented in section 4.

\subsection{Spin 1/2 fields}

Following Ref.~\cite{CL}, we define the incoming flux of a fermionic field as
the radial component of the conserved current, $J^\mu=\sqrt{2}\,\sigma^\mu_{AB}\,
\psi^A \bar\psi^B$, integrated over a two-dimensional sphere. This may be finally
written as
\beq
\frac{1}{2\pi}\,\frac{d N}{d t}=|R_{1/2}|^2 - |R_{-1/2}|^2\,,
\eeq
in terms of the measure squared of the upper and lower components of the
fermionic field. Taking the ratio of the value of this quantity at the horizon
with respect to its value at infinity, we obtain the absorption probability
for the scattering of an incoming wave travelling from spatial infinity towards
the horizon of the projected black hole. It can be simply expressed as
\beq
|A_{s=1/2}|^2=\biggl|\frac{A^{(h)}}{A^{(\infty)}}\biggl|^2
\eeq
if we take into account the fact that the upper component contributes little
to the infalling flux both at infinity and at the horizon.

Setting $s=1/2$ and expanding the $C$ and $D$ coefficients, defined in Eqs.
(\ref{C})-(\ref{D}), in the limit $\om r_H \ll 1$, we obtain
\bea
C C^* &=& \frac{2^{-(4j+2)/(n+1)}}{\Gamma(j+1)^2} + {\cal O}(\om r_H)^2\,, \\[3mm]
C D^* + C^* D &=& \frac{2^{2j+1}}{\pi} + {\cal O}(\om r_H)^2\,,\\[3mm]
D D^* &=& \frac{2^{4j}\,2^{(4j+2)/(n+1)}}{\pi^2}\,\Gamma(j+1)^2
+ {\cal O}(\om r_H)^2\,.
\eea
Taking into account the fact that the smallest physically allowed
value of the total angular
momentum $j$ is the value of the spin, $s=1/2$, we easily conclude that
the dominant term in the denominator of Eq. (\ref{ratio1}), in the low-energy
limit, is given by the first term in the expansion of $D D^*$. Moreover, we have
\beq
\frac{1}{|\Gamma(1/2+2 \alpha)|^2}=\frac{\cosh(2 \tilde\alpha \pi)}{\pi}
\simeq \frac{1}{\pi} + \frac{2\pi\,(\om r_H)^2}
{(n+1)^2} + {\cal O}(\om r_H)^3\,,
\label{sinh}
\eeq
where $\tilde \alpha$ is the imaginary part of $\alpha$. Putting 
everything together, we obtain
\beq
|A_{s=1/2}|^2 \simeq \frac{2\pi\,(\om r_H)^{2j+1}\,2^{-(4j+2)/(n+1)}}
{2^{2j}\,\Gamma(j+1)^2} + {\cal O}(\om r_H)^2\,.
\label{fermions}
\eeq

The absorption probability leads to the absorption cross section
$\si_{\rm abs}(\om)$ via
\beq 
\si^{(s)}_{\rm abs}(\om) = \frac{\pi\,(2j+1)}{\om^2}\,|A_{s}|^2
\label{sigmaabs}
\eeq
for any spin $s$ and angular momentum partial wave number $j$. Thus, for the
fermion case the final absorption cross section is, in leading order in
$(\om r_H)$,
\beq 
\si^{(1/2)}_{\rm abs}(\om) = \frac{2\pi^2\,r_H^2\,(2j+1)\,(\om r_H)^{2j-1}\,
2^{-(4j+2)/(n+1)}} {2^{2j}\,\Gamma(j+1)^2}\,.
\label{fermionsi}
\eeq
This is the primary analytic result valid for $\om r_H \ll 1$ 
for the case of spin 1/2 fermions.

As it is obvious from the above formula, the absorption cross-section has a
strong dependence on the total angular momentum number $j$ of the emitted
field as it affects both the power of the dominant $(\om r_H)$-term and the
multiplicative coefficient in front. On the other hand, as in the case of
the emission of brane-localised scalar fields studied in \cite{KMR1}, no
dependence of the power of $(\om r_H)$ on the number of extra dimensions
emerges for a brane-localised fermionic field. Nevertheless, the
multiplicative coefficient in front depends on the number of extra
dimensions.

In order to be able to draw concrete conclusions about the dependence
of $\sigma^{(1/2)}_{\rm abs}$ on $j$ and $n$, the value of the absorption
cross-section for $n=$ 2, 4 and 6, and $j=\frac{1}{2}$, $\frac{3}{2}$ and
$\frac{5}{2}$ have been computed and displayed in Table 1. By looking at the
entries of this Table for $j=\frac{1}{2}$ and for every $n$, we see that
the absorption cross section becomes proportional to the area of the horizon
of the black hole, times a numerical coefficient, and is independent of the
energy of the emitted particle.  As the total angular momentum number $j$
increases, the absorption cross section rapidly decreases as it gets
suppressed by extra powers of $\om r_H$ and its numerical coefficient
also gets smaller. If instead we fix $j$ and vary the number of extra
dimensions, then we see that the absorption cross section gets enhanced
as the number of the projected extra dimensions increases. 
The behaviour, therefore, of the leading term in the expression
of the greybody factor, in the low-energy regime, for the emission of
brane fermionic fields is similar to the one for the
emission of brane scalar fields \cite{KMR1}.

%
\begin{center}
$\begin{array}{|c||c||l|} \hline \hline 
\multicolumn{3}{|c|}{\rule[-3mm]{0mm}{8mm} 
{\bf Table~1: \rm \,\,Absorption\,\,Cross\,\,Section\,\,for\,\,a\,\,
brane-localised \,\,fermionic\,\,field}} \\ \hline\hline
{\rule[-3mm]{0mm}{8mm}
\hspace*{0.8cm}{\bf n=2} \hspace*{0.8cm}} & \hspace*{0.8cm} j=1/2 \hspace*{0.8cm}
& \hspace*{0.8cm} 
\si^{(1/2)}_{\rm abs} \simeq 2^{5/3}\,\pi\,r_H^2 +\ldots
\hspace*{0.8cm} \\[3mm]
 & j=3/2 &
\hspace*{0.8cm}  \si^{(1/2)}_{\rm abs} \simeq \frac{2^{4/3}}{9}\,
\pi\,r_H^2\,(\om r_H)^2 + \ldots \\[3mm]
 & j=5/2 &
\hspace*{0.8cm} \si^{(1/2)}_{\rm abs} \simeq \frac{\pi}{150}\,
r_H^2\,(\om r_H)^4 + \ldots
\\[3mm] \hline
{\rule[-3mm]{0mm}{8mm} {\bf n=4}  }& j=1/2 &
\hspace*{0.8cm} \si^{(1/2)}_{\rm abs} \simeq 4\,2^{1/5}\,\pi\,r_H^2+ \ldots
\\[3mm]
 & j=3/2 &
\hspace*{0.8cm} \si^{(1/2)}_{\rm abs} \simeq \frac{4\,\pi}{9}\,
2^{2/5}\,r_H^2\,(\om r_H)^2 + \dots
\\[3mm]
 & j=5/2 &
\hspace*{0.8cm} \si^{(1/2)}_{\rm abs} \simeq \frac{2^{3/5}}{75}\,
\pi\,r_H^2\,(\om r_H)^4 + \ldots
\\[3mm] \hline 
{\rule[-3mm]{0mm}{8mm} {\bf n=6} } & j=1/2 &
\hspace*{0.8cm} \si^{(1/2)}_{\rm abs} \simeq 4\,2^{3/7}\,
\pi\,r_H^2 + \ldots
\\[3mm]
 & j=3/2 &
\hspace*{0.8cm} \si^{(1/2)}_{\rm abs} \simeq \frac{4}{9}\,
2^{6/7}\,\pi\,r_H^2\,(\om r_H)^2 + \ldots
\\[3mm]
 & j=5/2 &
\hspace*{0.8cm} \si^{(1/2)}_{\rm abs} \simeq \frac{2}{75}\,
2^{2/7}\,\pi\,r_H^2\,(\om r_H)^4 + \ldots
\\[3mm] \hline \hline

\end{array}$
\end{center}

\subsection{Spin 1 fields}

We now move to the case of gauge fields. In this case, no conserved current may
be defined, however, the incoming flux can be expressed as the integral
of the flow of energy, $T^{\mu\nu}=2 \sigma^\mu_{AA'} \sigma^\nu_{BB'}
\psi^{AB} \bar\psi^{A'B'}$, through a 2-dimensional sphere \cite{CL}. Then
we obtain the expression,
\beq
\frac{1}{2\pi}\,\frac{d N}{d t}=\frac{1}{2\Sigma \om}\,\Bigl(|R_{1}|^2 -
|R_{-1}|^2\Bigr)\,,
\eeq
where the redshift of the local energy of photons has been taken into account.
The absorption probability is given again by the ratio of the value of the above
quantity at the horizon with respect to its value at infinity, and it is
found to be
\beq
|A_{s=1}|^2=\frac{1}{(2 \om r_H)^2}\,\biggl|\frac{A^{(h)}}{A^{(\infty)}}\biggl|^2\,.
\eeq

Going back to Eq. (\ref{ratio1}) and setting $s=1$, we expand the $C$ and $D$
coefficients in the limit $\om r_H \ll 1$. Thus, we obtain
\bea
C C^* &=& \frac{2^{4j+2}\,\Gamma\Bigl(-\frac{2j+1}{n+1}\Bigr)^2\,
\Gamma(j)^2}{\Gamma\Bigl(-\frac{j}{n+1}\Bigr)^2\,
\Gamma\Bigl(-\frac{j+1}{n+1}\Bigr)^2\,\Gamma(2j+1)^2}
+ {\cal O}(\om r_H)^2\,, \\[3mm]
C D^*  &=& \frac{-i\,2^{2j+1}}{\pi\,(n+1)}\,
\frac{\sin\bigl(\frac{\pi j}{n+1}\Bigr)\,
\sin\Bigl(\frac{\pi\,(j+1)}{n+1}\Bigr)}
{\sin\Bigl(\frac{\pi\,(2j+1)}{n+1}\Bigr)} + {\cal O}(\om r_H)\,,\\[3mm]
D D^* &=& \frac{\Gamma\Bigl(\frac{2j+1}{n+1}\Bigr)^2\,
\Gamma(2j+2)^2}{\Gamma\Bigl(\frac{j}{n+1}\Bigr)^2\,
\Gamma\Bigl(\frac{j+1}{n+1}\Bigr)^2\,\Gamma(j+2)^2}
+ {\cal O}(\om r_H)^2\,.
\eea
The first term in the expansion of $D D^*$ gives again the dominant term in
the denominator of Eq. (\ref{ratio1}) in the low-energy limit. Also, we have
\beq
\frac{1}{\Ga(2\alpha)\,\Ga(-2\alpha)}=\frac{2 \tilde \alpha \,
\sinh(2 \tilde \alpha \pi)}{\pi} \simeq \frac{4\,(\om r_H)^2}
{(n+1)^2} + {\cal O}(\om r_H)^4\,.
\label{sinh-g}
\eeq
By using the above results, the absorption probability may be written as 
\beq
|A_{s=1}|^2\simeq
\frac{4(\om r_H)^2 (2 \om r_H)^{2j}}{(n+1)^2}\,
\frac{\Gamma\Bigl(\frac{j}{n+1}\Bigr)^2\,
\Gamma\Bigl(\frac{j+1}{n+1}\Bigr)^2\,\Gamma(j+2)^2}
{\Gamma\Bigl(\frac{2j+1}{n+1}\Bigr)^2\,\Gamma(2j+2)^2} +
{\cal O}(\om r_H)^2\,,
\label{bosons}
\eeq
leading to the following absorption cross section, for the gauge field case,
in leading order in $\om r_H$,
\beq 
\si^{(1)}_{\rm abs}(\om) = 4\pi r_H^2 \,(2\om r_H)^{2j}\,\frac{(2j+1)}{(n+1)^2}\,
\left(\frac{\Gamma\Bigl(\frac{j}{n+1}\Bigr)\,
\Gamma\Bigl(\frac{j+1}{n+1}\Bigr)\,\Gamma(j+2)}
{\Gamma\Bigl(\frac{2j+1}{n+1}\Bigr)\,\Gamma(2j+2)}\right)^2\,.
\label{gaugesi}
\eeq
This is our primary analytic result valid for $\om r_H\ll 1$ 
for the case of gauge fields.

As in the case of scalar and fermionic fields, the absorption cross section for
brane localised gauge fields depends non-trivially on the angular momentum
number $j$ and the number of the extra dimensions $n$. The power of the
dominant $(\om r_H)$-term depends only on $j$ but the numerical coefficient
in front depends strongly on both parameters. The values of the above quantity
for $n=$2, 4 and 6, and $j=$1, 2 and 3 are given in Table 2.

%
\begin{center}
$\begin{array}{|c||c||l|} \hline \hline 
\multicolumn{3}{|c|}{\rule[-3mm]{0mm}{8mm} 
{\bf Table~2: \rm \,\,Absorption\,\,Cross\,\,Section\,\,for\,\,a\,\,
brane-localised \,\,gauge\,\,field}} \\ \hline\hline
{\rule[-3mm]{0mm}{8mm}
\hspace*{0.8cm}{\bf n=2} \hspace*{0.8cm}} & \hspace*{0.8cm} j=1 \hspace*{0.8cm}
& \hspace*{0.8cm} 
\si^{(1)}_{\rm abs} \simeq \frac{64}{81}\,\pi^3\,r_H^2\,(\om r_H)^2 +\ldots \hspace*{0.8cm}
\\[3mm]
 & j=2 &
\hspace*{0.8cm}  \si^{(1)}_{\rm abs} \simeq \frac{\pi}{5}\,
r_H^2\,(\om r_H)^4 + \ldots \\[3mm]
 & j=3 &
\hspace*{0.8cm} \si^{(1)}_{\rm abs} \simeq \frac{4\pi}{1575}\,
r_H^2\,(\om r_H)^6 + \ldots
\\[3mm] \hline
{\rule[-3mm]{0mm}{8mm} {\bf n=4}  }& j=1 &
\hspace*{0.8cm} \si^{(1)}_{\rm abs} \simeq \frac{4}{75}\,2^{2/5}\,
\Gamma(\frac{1}{10})^2\,\Gamma(\frac{2}{5})^2\,r_H^2\,(\om r_H)^2 + \ldots
\\[3mm]
 & j=2 &
\hspace*{0.8cm} \si^{(1)}_{\rm abs} \simeq \frac{4\pi}{125}\,
\Ga(\frac{2}{5})^2\,\Ga(\frac{3}{5})^2\,r_H^2\,(\om r_H)^4 + \dots
\\[3mm]
 & j=3 &
\hspace*{0.8cm} \si^{(1)}_{\rm abs} \simeq \frac{64 \pi}{39375}\,
\frac{\Ga(3/5)^2\,\Ga(4/5)^2}{\Ga(7/5)^2}\,r_H^2\,(\om r_H)^6 + \ldots
\\[3mm] \hline 
{\rule[-3mm]{0mm}{8mm} {\bf n=6} } & j=1 &
\hspace*{0.8cm} \si^{(1)}_{\rm abs} \simeq \frac{16 \pi}{147}\,
\frac{\Ga(1/7)^2\,\Ga(2/7)^2}{\Ga(3/7)^2}\,r_H^2\,(\om r_H)^2 + \ldots
\\[3mm]
 & j=2 &
\hspace*{0.8cm} \si^{(1)}_{\rm abs} \simeq \frac{2^{6/7}}{245}\,
\Ga(\frac{3}{14})^2\,\Ga(\frac{2}{7})^2\,r_H^2\,(\om r_H)^4 + \ldots
\\[3mm]
 & j=3 &
\hspace*{0.8cm} \si^{(1)}_{\rm abs} \simeq \frac{64}{77175}\,
\frac{\pi^3}{\sin^2(4 \pi/7)}\,r_H^2\,(\om r_H)^6 + \ldots
\\[3mm] \hline \hline
\end{array}$
\end{center}

Unlike the case of scalar and fermionic field emission, the absorption
cross section for the emission of gauge fields depends on the energy of
the emitted particle even in the case of the lowest partial wave with $j=1$.
This follows from the different dependence of the dominant $(\om r_H)$-term
on the angular momentum number. Any further increase in the value of
$j$ leads to the appearance of a suppression factor of powers of $(\om r_H)$
while the increase in the number of extra dimensions strongly enhances the
absorption cross section.

\section{Evaluation of Results}

In this section, we proceed to calculate the corresponding emission
rates for spin 1/2 and 1 fields in the background of the projected
4-dimensional black hole given in Eq. (\ref{projected}).  In this
section we will not use the simplified expressions valid only
for $\om r_H \ll 1$, but rather the full analytic expressions
Eqs.~(36-38). 

For the relevant 4-dimensional process of the emission
of brane-localised fields, Eq.~(\ref{eq:greybody}) takes the form
\beq
\frac{dE(\om)}{dt} = \sum_{j} \sigma_{j}(\om)\,{\om^3  \over
\exp\left(\om/T_{BH}\right) \mp 1}\,\frac{d \omega}{2\pi^2}\,.
\label{emission}
\eeq
Thus for the evaluation of the emission rates, we need the expressions of
the greybody factors summed over the angular momentum number $j$.
A simple numerical analysis shows that, by summing over the first three
partial waves, we obtain the dependence of the greybody factors
on the parameter $\omega r_H$ to high accuracy, with
all higher partial waves adding an almost zero contribution
to the final result. The numerical
evaluation gives us the ability to use the exact value of the ratio
(\ref{ratio1}) instead of the simplified leading-order corrections
presented in Eqs. (\ref{fermionsi}) and (\ref{gaugesi}), which are
valid only for $\omega r_H \ll 1$. Such a numerical analysis will
reveal how fast the low-energy approximation breaks down as the
expansion parameter $\omega r_H$ increases. Figures 1(a) and (b) depict
the greybody factors for fermions and gauge fields, respectively,
in units of $\pi r_H^2$, as functions of $\omega r_H$. 
The first three partial waves have been summed in each case
and the dependence for $n=0,2,4$ and 6 is shown. We observe that, for
small enough values of $\omega r_H$, we reproduce, as expected, the
behaviour given by the leading-order corrections and according to
which the greybody factors are enhanced as the number of extra dimensions
projected on our brane increases. This agreement holds for both 
species of fields up to the value $\omega r_H\simeq 0.4$. As
$\omega r_H$ increases further, it is the lower-dimensional models
that seem to give the largest greybody factors. However, our
calculation is limited to a WKB-like semiclassical approximation 
and only an exact numerical analysis of the original master equation,
Eq. (11), can determine $\sigma_{\rm abs}(\omega)$ in the
intermediate and high-energy regime $\om r_H \gsim 1$. Let
us finally note that our analysis correctly reproduces
the behaviour of the greybody factors in the case of $n=0$ \cite{classics,MW},
with $\sigma^{(1/2)}_{\rm abs}(\omega)$ adopting a non-vanishing asymptotic
value at very low energies and $\sigma^{(1)}_{\rm abs}(\omega)$ going to
zero. The same behaviour is observed in the cases with $n \neq 0$.
%
\begin{center}
\begin{figure}[t]
\centerline{\hspace*{-1cm}\hbox{\psfig{file=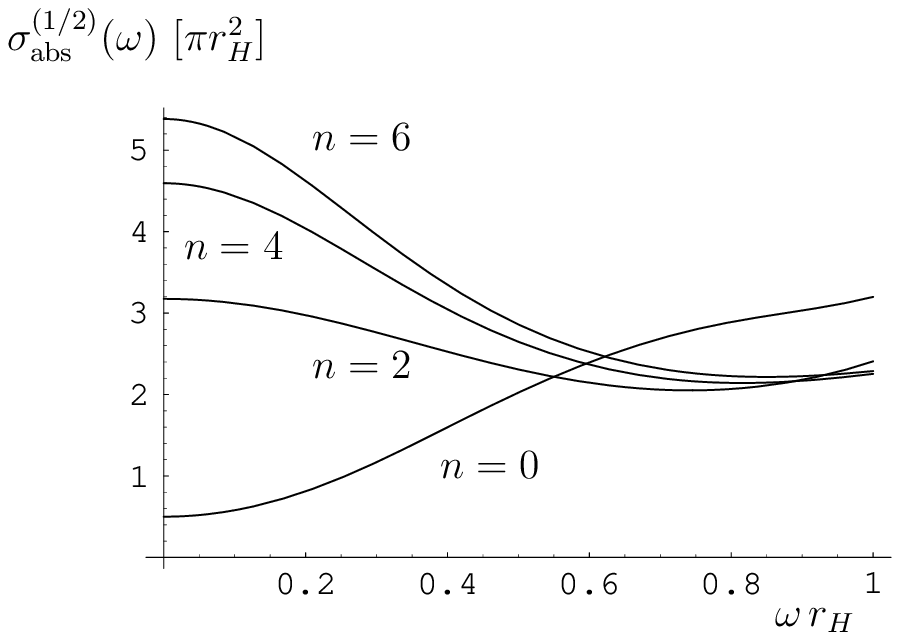, height=5cm}}
\hspace*{0.5cm}
{\psfig{file=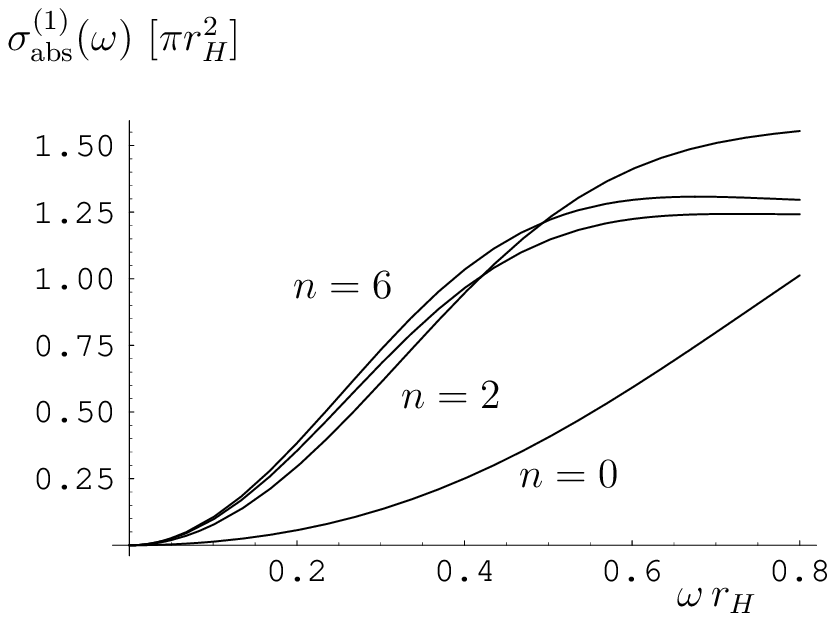, height=5cm}}}
\caption{\it The greybody factors $\sigma_{\rm abs}(\omega)$ for the emission
of {\bf (a)} fermions, and
{\bf (b)} gauge brane-localized fields, as a function of $\omega r_H$,
for $n=0,2,4$ and 6.}
\end{figure}
\end{center}
\vspace*{-1.5cm}

Figures 2(a) and (b) display the behaviour of the differential emission rate
per time unit and energy interval $d^2E/dt\,d\omega$ versus the parameter
$\omega r_H$, for fermions and gauge fields, as it follows by combining
Eq. (\ref{emission})
with the exact expression of the greybody factor in each case and the
definition of the temperature of the projected 4-dimensional black hole,
$T_H=(n+1)/4 \pi r_H$. The emission rates are substantially enhanced as
the number of extra dimensions increases. The increase is larger for
gauge fields which leads to the result that the emission
rates for the two species become comparable, for large values of $n$.
On the other hand, for low values of $n$, the emission rate for a
fermion is considerably larger compared than that for a gauge field,
with the difference reaching an order of magnitude
in the limiting case of $n=0$. 

It is useful to perform a similar numerical analysis for the
case of scalar fields being emitted by the same 4-dimensional black
hole, a case studied in Ref. \cite{KMR1}. By following a similar
analytical approach, the leading-order correction in the expression
of the greybody factor in that case was also derived. According to
those results, the leading correction for the first, and thus dominant,
partial wave was independent of the number of extra dimensions. Any
difference in the value of the greybody factor was to come from the
higher partial waves whose coefficient indeed increased with increasing
$n$. As $\omega r_H$ increases, however, the next-to-leading-order corrections
for each partial wave (for example, the terms being denoted by ``..." in
Table 1 and 2, and similarly in Table II in \cite{KMR1}) can dominate
the contribution coming from the higher partial
waves. In the case of 
%
\begin{center}
\begin{figure}[t]
\centerline{\hspace*{-1cm}\hbox{\psfig{file=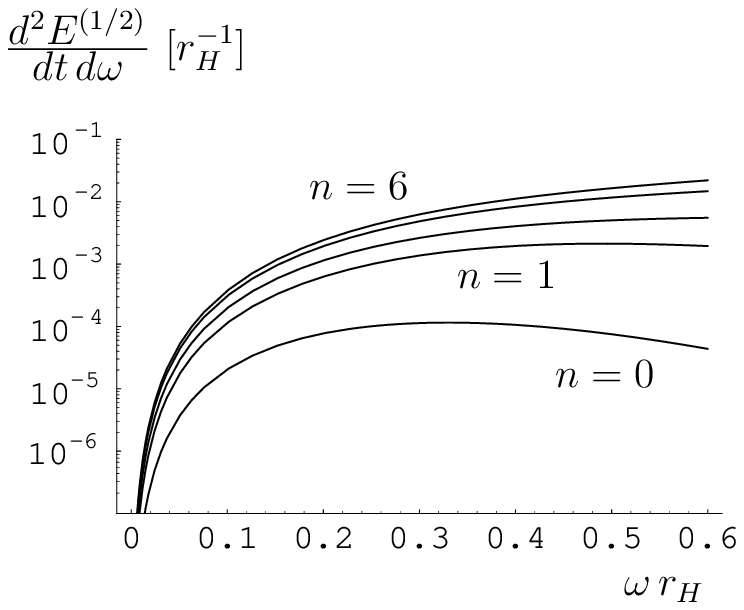, height=5cm}}
\hspace*{0.5cm}
{\psfig{file=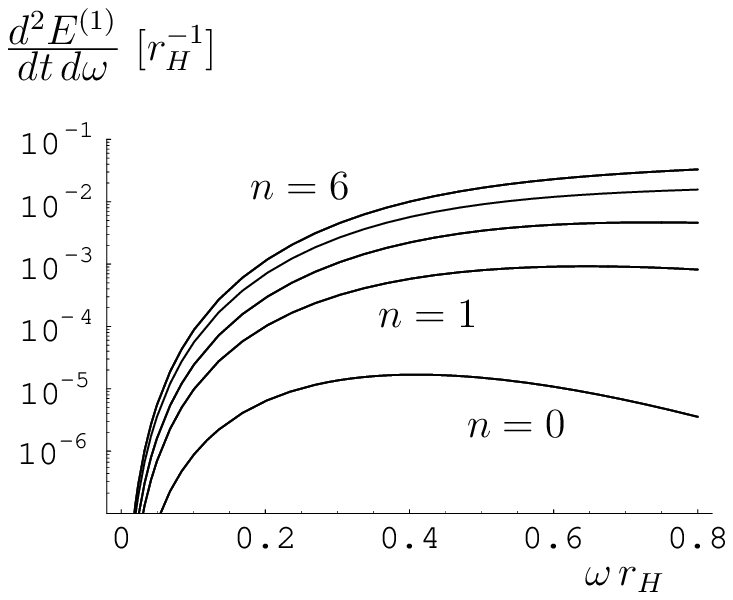, height=5cm}}}
\caption{\it The differential emission rate $d^2 E/dt\,d\omega$ 
for {\bf (a)} fermions and {\bf (b)} gauge fields, as a function of
$\omega r_H$, for $n=0,1,2,4$ and 6.}
\end{figure}
\end{center}
\vspace*{-1.3cm}
%
\begin{center}
\vspace*{-1.25cm}
\begin{figure}[b]
\centerline{\hspace*{-1cm}\hbox{\psfig{file=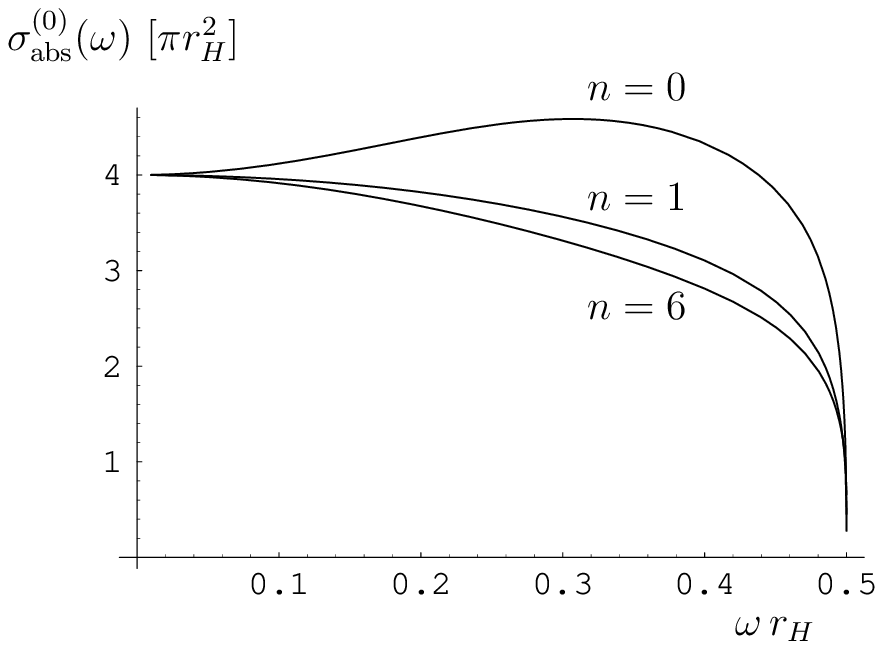, height=5cm}}
\hspace*{0.5cm}
{\psfig{file=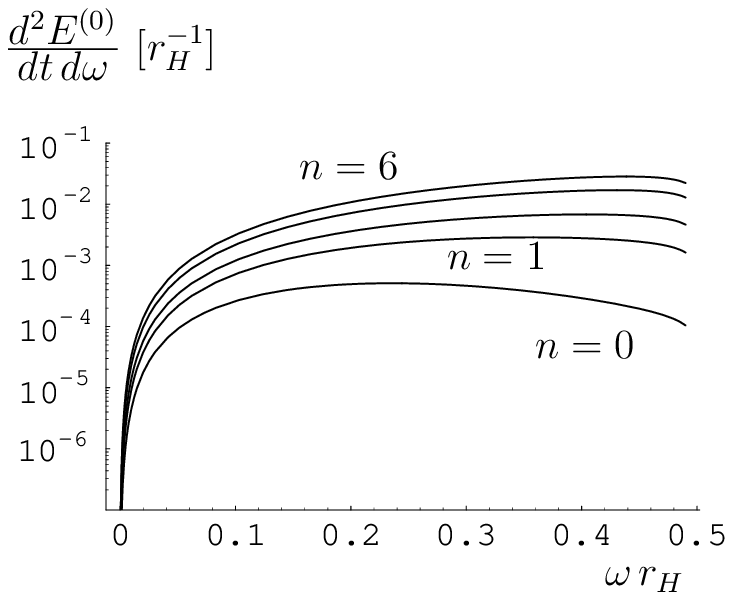, height=5cm}}}
\caption{\it {\bf (a)} The greybody factor for the emission of brane-localized
scalar fields for $n=0,1$ and 6; {\bf (b)} the corresponding differential
emission rate for $n=0,1,2,4$ and 6.}
\end{figure}
\end{center}
%

\noindent
fermions and gauge fields, we showed that this
does not happen until the value $\omega r_H \simeq 0.4$. For scalar
fields, it turns out that this value is much smaller. Figure 3(a)
depicts the greybody factor for the emission of brane-localized
scalar fields, by using the exact value of the ratio of the two
coefficients $B_+$ and $B_-$, given in Eq. (46) of
Ref. \cite{KMR1}. We immediately see that, for extremely small values
of $\omega r_H$, the greybody factors, for various values of $n$,
converge to the same asymptotic value, in agreement with our simplified
expression. However, very soon, the next-to-leading-order corrections
dominate and cause the greybody factor to decrease as $n$ increases. 
The emission rate for scalar fields is nevertheless enhanced as the
number of extra dimensions
increases. This is caused by the decrease in the denominator of
Eq. (\ref{emission}), due to the dependence of the temperature $T_{BH}$
on $n$, that overcomes the decrease of the greybody factor. As in the
case of fermion and gauge fields, the increase amounts to orders of
magnitude compared to the limiting case of $n=0$.

\section{Conclusions}

In an earlier paper \cite{KMR1}, we studied the problem of the decay of a 
higher-dimensional Schwarzschild-like black hole through the emission of
either higher-dimensional bulk, or brane-localised, scalar modes.
In this paper, we extended the computation of the greybody factors
that appear in the Hawking decay process to the spin 1/2 and spin 1
brane-localised cases of greatest phenomenological interest.
Since the angular momentum of the black hole has been ignored, our
calculations are relevant to the post balding and spin-down phases of
the life of a black hole produced at a high energy collider or by
ultra high energy cosmic ray interactions.  The semiclassical approximation
methods employed allow us to calculate the greybody factors and the
corresponding emission rates for a general number $n$ of flat
extra dimensions of the Arkani-Hamed,
Dimopoulos, and Dvali type.  (It is important to bear in mind that black
holes in highly curved bulk spacetimes, for example Randall-Sundrum 5d
theories, require a somewhat modified treatment from the one presented
here.)  The primary analytic results of our calculation for the
greybody factors, for arbitrary $j$ and $n$, are presented in
Eqs.~(35-38), and (40), (46), and (49).  Simplified expressions
are provided in Eqs.~(47) and (55) spin 1/2 and spin 1 respectively.
In both cases there is an increase in the greybody factor as $n$ is
increased at fixed $j$. As the subsequent numerical analysis revealed,
this behaviour survives up to intermediate values of the energy of the
emitted particle and contributes to a considerably enhanced emission
rate for both fermions and gauge bosons. A similar analysis for the emission
of scalar fields also led to an enhancement of the emission rate as
$n$ increases despite the fact that the greybody factor, rapidly deviating
from the behaviour dictated by the leading-order correction in the
low-energy approximation, is suppressed.
We expect that our results will be of use in the detailed study of the
signatures of possible black hole production events.

\noindent
{\bf Note added}.
While writing this paper, the related work of Ref. \cite{park} appeared
which also considers the greybody factors for black holes in brane-world
theories.

\noindent
{\bf Acknowledgments.} 
We wish to thank Roberto Emparan and Nemanja Kaloper for discussions.

\end{document}